# Revisiting the Voltage-Source Behavior: Why Impedance Magnitude of Grid-Forming Converter Rises Near Fundamental Frequency?

Chao Wu, *Senior Member*, *IEEE*, Jinhao Wang, Yong Wang, *Member*, *IEEE*, Changjiang Zhan, *Senior Member*, *IEEE*, and Frede Blaabjerg, *Fellow, IEEE*

***Abstract*—Grid-forming (GFM) converters are generally expected to exhibit low impedance near the fundamental frequency due to their voltage-source behavior. However, an impedance peak and a negative-resistance region are consistently observed in this range, which contradicts this expectation and lacks a clear physical explanation. This paper reveals that these phenomena originate from the inherent dynamics of the active power control loop, where the mapping from power disturbance to the synchronous angle inherently involves an integrative action, intrinsically preventing a positive-resistance characteristic near the fundamental frequency. This finding explains why existing grid codes in China, the United States, and Europe exclude a narrow band around the fundamental frequency in impedance-based evaluations. It is further shown that the width of the excluded frequency band (e.g., ±3–5 Hz) is governed by the power-to-frequency dynamics. Based on this insight, a quantitative index is proposed to determine the exclusion bandwidth from the corner frequencies of the impedance magnitude curve. The proposed index provides a concise and theoretically grounded criterion for voltage-source assessment and impedance standardization of GFM converters.**

***Index Terms*—Grid-forming (GFM) converters, voltage-source behavior, impedance peak, negative-resistance region, active power control loop (APCL).**

## I. INTRODUCTION

Grid-forming (GFM) converters have attracted significant attention for enabling stable operation in power-electronics-dominated systems due to their capability to support voltage and frequency [1]. In recent years, extensive studies have been conducted on modeling [2], stability analysis [3], and control optimization of GFM converters [4]. In contrast to state-space modeling, which is typically regarded as a white-box approach based on eigenvalue analysis and requires detailed knowledge of system structure and parameters [5], impedance-based modeling can be treated as a black-box approach that does not rely on complete design information [6]. Instead, impedance characteristics can be directly obtained through measurement, enabling effective validation and making this approach particularly suitable for investigating oscillation mechanisms in grid-connected GFM systems [7], [8]. Nevertheless, existing research has predominantly focused on stability assessment and corresponding control-oriented optimization, while the fundamental voltage-source behavior exhibited by GFM converters have received comparatively limited attention [9], [10].

Traditionally, it is expected that a voltage source should exhibit low impedance magnitudes near the fundamental frequency, as governed by the bandwidth of the voltage control loop. For instance, the *Great Britain Grid Forming (GBGF) Best Practice Guide* and the *Australian Energy Market Operator (AEMO)* both state that the *GFM converter should exhibit a low impedance magnitude around the fundamental frequency, due to its voltage-source behavior* [11], [12]. However, impedance of the GFM converter shows that the low impedance occurs around 35-40 Hz, followed by a rising trend toward 50 Hz [13], [14]. Contrary to the conventional expectation, a distinct magnitude peak is observed at the fundamental frequency, which challenges this widely held consensus.

Recent studies further confirm that this impedance peak is not an exception. In GFM converters equipped with inner-loop control structures, the positive-sequence impedance exhibits a distinct magnitude peak near the fundamental frequency [15]. Interestingly, the same peak is also observed in converters without inner-loop control [16], indicating that this behavior persists regardless of the presence or absence of an inner-loop framework. Moreover, as shown in [17], the impedance peak near the fundamental frequency consistently appears across various control configurations, including voltage–current dual-loop control, voltage single-loop control, and inner-loop-free control, indicating that its occurrence is independent of the specific inner-loop control strategy. This behavior has been attributed to droop-based power control [18]; however, it remains unclear whether active or reactive power droop dominates, and more fundamentally, which component of the power control loop is primarily responsible. Collectively, these observations indicate that the impedance peak near the fundamental frequency is an inherent characteristic, underscoring the necessity of a mechanistic framework to rigorously identify its root cause.

Beyond the theoretical interest, the phenomenon also carries significant implications for practical engineering applications and standardization efforts. From an engineering application perspective, understanding the impedance peak near the fundamental frequency is of great importance for evaluating the voltage-source behavior of GFM converters and for guiding the development of impedance standards. In 2023, the *North American Electric Reliability Corporation (NERC)* specified that *GFM should present a non-negative resistance to the grid within a wide frequency range (0-300 Hz)* [19]. However, in the same year, *FINGRID* required that *GFM shall present a positive resistance to the grid within frequency ranges 0–47 Hz and 53-250 Hz* [20], explicitly excluding the ±3 Hz region around the fundamental frequency. Similarly, in China, the *Technical Specification for Grid-Forming Power Conversion System of Electrochemical Energy Storage (Draft for Comments)* requires that *GFM converters shall present positive resistance within 1–45 Hz and 55–1000 Hz* [21], thereby excluding the ±5 Hz band around the fundamental frequency. More recently, the UNIFI Consortium in the United States also states that *GFM resource should have a positive-sequence impedance response similar to that of an R-L branch within* $f_1$ *± 40 Hz frequency range except for a narrow band around the fundamental frequency* ($f_1$*=60 Hz*), for which a recommended exclusion range of ±4 Hz is provided [22], [23]. Despite these efforts, existing grid codes and technical reports generally lack a clear physical explanation for excluding the fundamental-frequency region and do not provide a rigorous or unified criterion for determining the exclusion bandwidth.

Overall, the impedance peak and negative-resistance region of GFM converters near the fundamental frequency remain insufficiently explained and appear to contradict conventional understanding, leaving their underlying mechanism unclear. To address this gap, this paper investigates the fundamental mechanism of these phenomena and identifies the key contributing factors. The main contributions and innovations of this paper are summarized as follows.

1) An analytical model is proposed to reveal that the impedance peak and negative-resistance region near the fundamental frequency in GFM converters originate from the mapping of power disturbances to the synchronous angle, inherently introduced by the active power control loop (APCL). The model further explains why a narrow frequency band around the fundamental frequency must be excluded in impedance-based evaluations and why a fully positive-resistance characteristic cannot be achieved over the entire frequency range.
2) It is further demonstrated that the frequency range of the impedance peak and the associated negative-resistance region is quantitatively determined by the APCL parameters through the power-to-frequency control function. Specifically, larger inertia and damping coefficients lead to a narrower affected frequency range, whereas the frequency-to-angle integrator in APCL determines the resulting peak magnitude.
3) A quantitative index is proposed based on the corner frequencies of the impedance magnitude curve, identified through the sign change of its slope around the fundamental frequency. The resulting symmetric exclusion bandwidth $\pm\Delta f$ provides a concise and theoretically grounded criterion for voltage-source assessment, offering practical guidance for impedance testing and future standardization of GFM converters.

The remainder of this paper is organized as follows. Section II analyzes the impact of sequentially added control loops on the impedance characteristics. Section III derives an analytical model to explain the impedance peak mechanism. Section IV proposes a new index to exclude the impedance peak and negative-resistance region. Section V presents experimental validation, and Section VI concludes the paper.

## II. Impedance Characteristic of GFM Converters Shaped by Sequential Control Loop Integration

The impedance characteristics of GFM converters are analyzed by examining how control loops shape their behavior near the fundamental frequency. This section develops impedance models by sequentially introducing current, voltage, and power control loops, revealing the dynamic influence of each on the converter's impedance characteristics. The studied GFM converter is illustrated in Fig. 1. Here, $\boldsymbol{V}$ denotes the point of common coupling (PCC) voltage, $\boldsymbol{I}$ represents the converter output current, and $V_{dc}$ is the dc-link voltage. $L_g$ and $R_g$ denote the grid inductance and resistance, respectively, while $L_f$ is the filter inductance. It is worth noting that the output filter is typically an LC-filter, but the effect of the capacitor near the fundamental frequency is negligible [24].

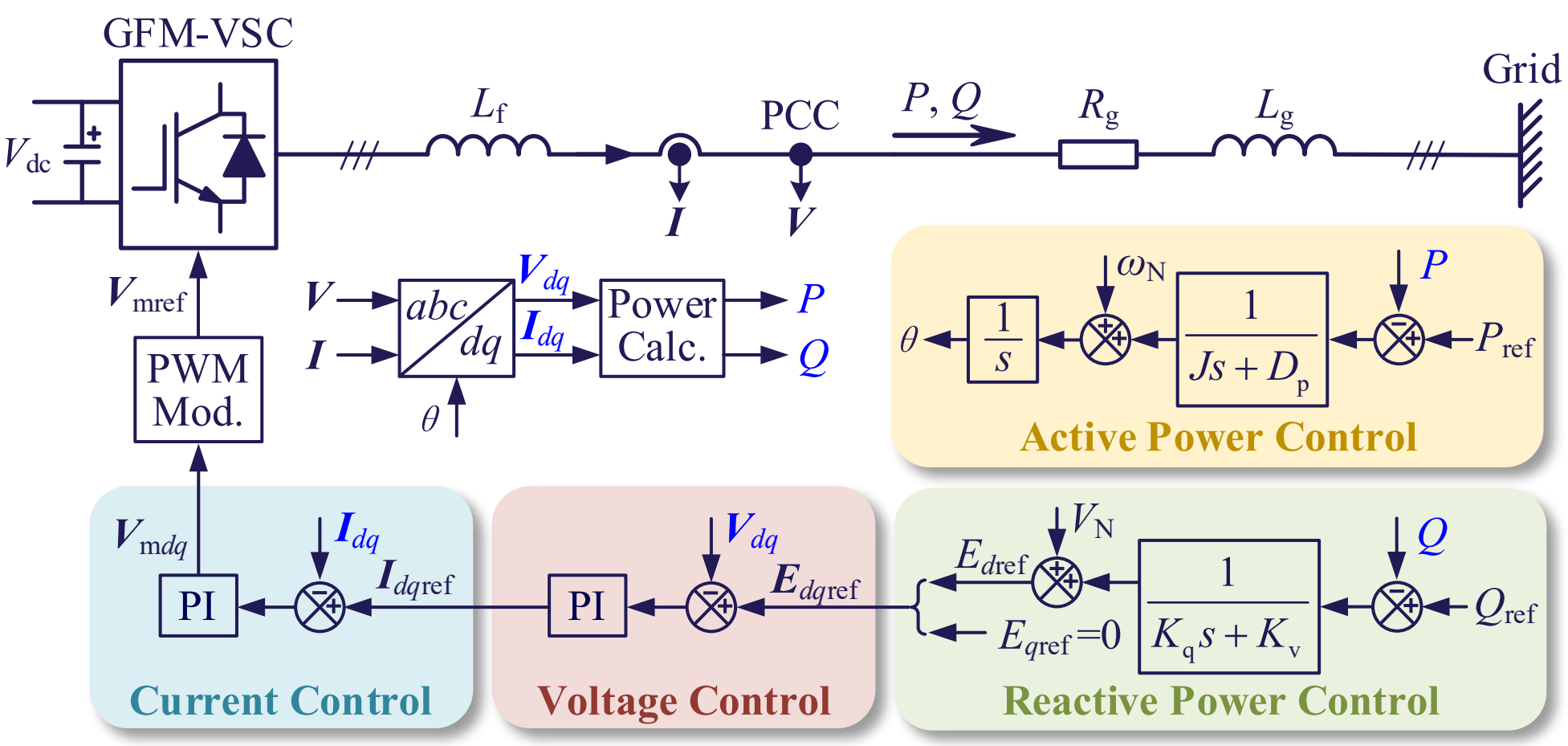

**Fig. 1.** Structure of the GFM converter connected to the grid.

TABLE I
PARAMETERS OF GFM CONVERTER

| Parameter | Value | Parameter | Value |
|---|---|---|---|
| $S_N$ | 200 kVA (1.0 p.u.) | $\omega_N$ | $100\pi$ rad/s (1.0 p.u.) |
| $V_N$ | 563 V (1.0 p.u.) | $J$ | 2546 ($s^2$·W)/rad (4 p.u.) |
| $I_N$ | 236 A (1.0 p.u.) | $D_p$ | 31832 (s·W)/rad (50 p.u.) |
| $V_{dc}$ | 1300 V | $K_v$ | 4438 Var/V (12.5 p.u.) |
| $P_{ref}$ | 200 kW (1.0 p.u.) | $K_q$ | 0.01 (s·V)/Var (4.0 p.u.) |
| $Q_{ref}$ | 0 Var | $k_{pV}$, $k_{iV}$ | 0.04 S, 347 S/s |
| $L_f$ | 300 μH (0.04 p.u.) | $k_{pI}$, $k_{iI}$ | 1.26 Ω, 420Ω/s |

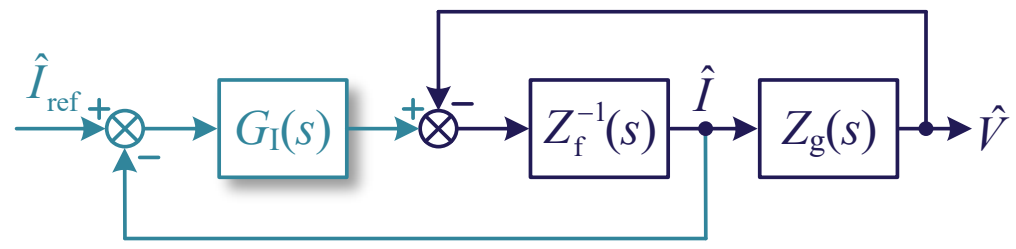

**Fig. 2.** Small-signal control block diagram with current loop only.

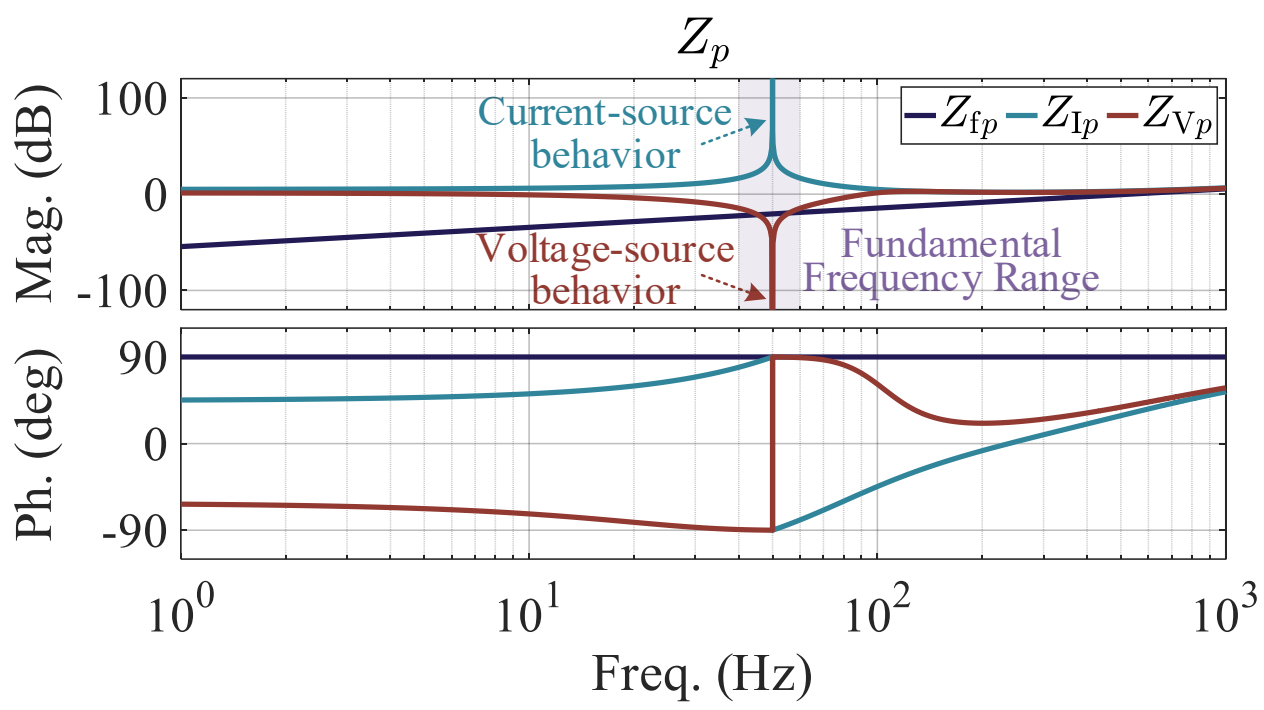

**Fig. 3.** Positive-sequence impedance of GFM converter with inner loops only.

$P_{ref}$ and $Q_{ref}$ are the active and reactive power references, with $P$ and $Q$ denoting their actual values flowing from the converter side to the grid side. $J$ is the inertia constant, $D_p$ is the damping coefficient, $K_v$ is the voltage regulation gain, and $K_q$ is the integral gain. $V_N$ and $\omega_N$ denote the rated voltage and angular frequency, respectively. $\theta$ represents the synchronization angle generated by the active power control, and both $\boldsymbol{V}$ and $\boldsymbol{I}$ are expressed in terms of their $d$- and $q$-axis components in the synchronous rotating frame defined by $\theta$. In addition, since this work mainly investigates impedance characteristics near the fundamental frequency, high-frequency dynamics caused by time delays are not considered [25]. The studied converter is a 690 V/200 kW commercial GFM unit, with its detailed parameters summarized in Table I.

*A. Modeling and Analysis of GFM Converter with Inner Loop*

The modeling begins with the current control loop (CCL). It is noted that when the outer power control loops are neglected, the system can be represented by a single-input single-output (SISO) transfer function. Fig. 2 shows the small-signal block diagram considering only current loop dynamics, from which the corresponding impedance $Z_{Idq}$ is derived as:

$$Z_{Idq} = \frac{\hat{V}}{-\hat{I}} = Z_f + G_I = sL_f + k_{pI} + \frac{k_{iI}}{s} \tag{1}$$

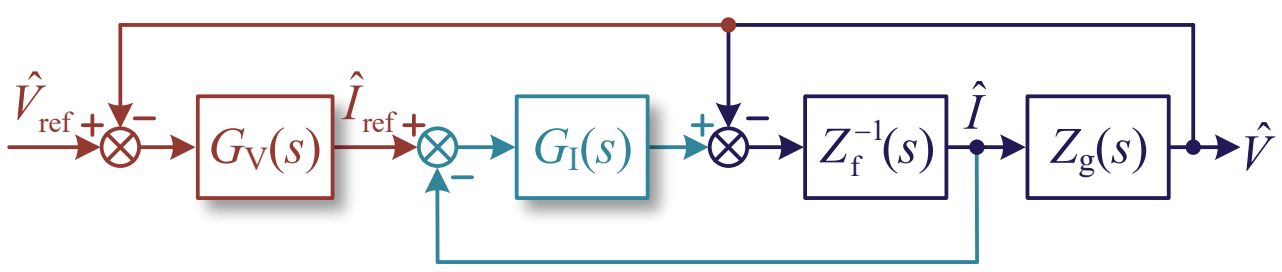

**Fig. 4.** Small-signal control block diagram with current and voltage loops.

Here, $Z_f$ represents the transfer function of the filter inductor, and $G_I$ denotes the transfer function of the current controller. The superscript "^" denotes small-signal perturbation. The subscript "$_{dq}$" denotes the scalar impedance in the rotating reference frame, whereas "$_p$" indicates the positive-sequence impedance in the stationary reference frame, with the corresponding transformations detailed in [26] and adopted throughout the remainder of this paper.

From (1), the CCL effectively adds a control term in series with the filter impedance $Z_{fp}$. Its positive-sequence impedance $Z_{Ip}$ in the stationary frame is plotted in Fig. 3, where the integrator leads to a high magnitude near the fundamental frequency, indicating a current-source behavior.

Next, the voltage control loop (VCL) is incorporated. Fig. 4 shows the small-signal control block diagram with both current and voltage loops. Based on Fig. 4, the corresponding impedance model $Z_{Vdq}$ is derived as:

$$\begin{aligned} Z_{Vdq} &= \frac{\hat{V}}{-\hat{I}} = \frac{Z_f + G_I}{1+G_V G_I} = \frac{Z_f}{1+G_V G_I} + G_I // \frac{1}{G_V} \\ &= \frac{s\left(s^2 L_f + s k_{pI} + k_{iI}\right)}{s^2\left(k_{pI}k_{pV}+1\right) + s\left(k_{pI}k_{iV} + k_{iI}k_{pV}\right) + k_{iI}k_{iV}} \end{aligned} \tag{2}$$

where $G_V = k_{pV} + \frac{k_{iV}}{s}$ denotes the transfer function of the voltage controller.

As seen from (2), the VCL introduces a parallel control path by adding the inverse of the voltage controller $1/G_V$ in parallel with the current controller $G_I$, while also reshaping the original filter impedance $Z_f$. The resulting positive-sequence impedance $Z_{Vp}$ is also shown in Fig. 3. Due to the differentiating action in (2), the impedance magnitude near the fundamental frequency is significantly reduced, indicating a voltage-source behavior. It is worth noting that, from the phase-response perspective, when only the voltage and current inner-loop dynamics are considered, the GFM converter does not exhibit a negative-resistance region and thus maintains a positive-resistance over the entire frequency range.

*B. Modeling and Analysis of GFM Converter with Outer Loop*

The analysis then examines the outer power control loop, starting with the reactive power control loop (RPCL). It is noted that when the power control loops are incorporated, the system must be described by a two-by-two multiple-input multiple-output (MIMO) impedance matrix. The modeling of the GFM converter with RPCL, VCL, and CCL has been presented in previous work [7], so it is not repeated here. Instead, the small-signal control block diagram is directly shown in Fig. 5, and the corresponding impedance model $\boldsymbol{Z}_{Qdq}$ is given as:

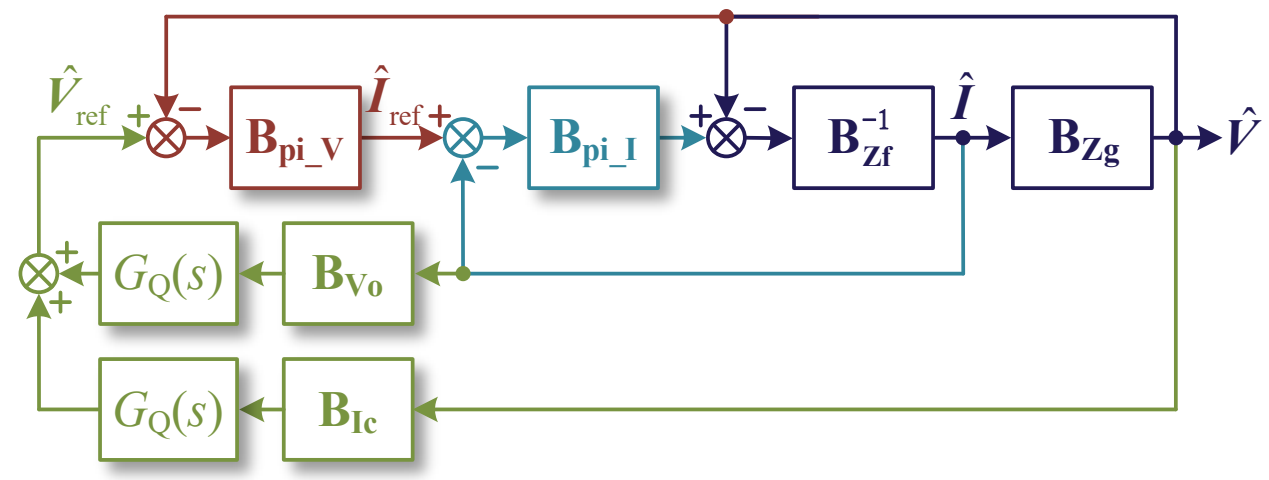

**Fig. 5.** Small-signal control block diagram with RPCL and inner loops.

$$\boldsymbol{Z}_{Qdq}=\begin{bmatrix}\dfrac{Z_f+G_I}{1+G_VG_I} & Z_{Qdq12}\\ 0 & \dfrac{Z_f+G_I}{1+G_VG_I}\end{bmatrix} \tag{3}$$

where $Z_{Qdq12}=\dfrac{G_QG_VG_I\left[v_{d0}\left(1+G_VG_I\right)+i_{d0}\left(Z_f+G_I\right)\right]}{\left(1+G_VG_I\right)^2}$, and $G_Q=\dfrac{1}{K_qs+K_v}$ denotes the transfer function of the reactive power controller, while $v_{d0}$ and $i_{d0}$ represent the steady-state $d$-axis components of the PCC voltage and converter current, respectively.

Fig. 6 shows the related positive-sequence impedance $Z_{Qp}$. By including the RPCL dynamics, it will result in a low impedance magnitude in the fundamental frequency range, and the converter still behaves as a voltage source. Therefore, RPCL is not the root cause of the impedance rise, and its dynamics are omitted later.

Finally, the active power control loop (APCL) is discussed, with its small-signal control block diagram presented in Fig. 7. The steady-state operating point matrices introduced by the active power control dynamics are expressed as:

$$\begin{cases}\mathbf{B_{Vo_i}}=\begin{bmatrix}-v_{q0}i_{d0} & -v_{q0}i_{q0}\\ v_{d0}i_{d0} & v_{d0}i_{q0}\end{bmatrix}, \mathbf{B_{Vo_v}}=\begin{bmatrix}-v_{d0}v_{q0} & -v_{q0}^2\\ v_{d0}^2 & v_{d0}v_{q0}\end{bmatrix}\\ \mathbf{B_{Ic_i}}=\begin{bmatrix}-i_{d0}i_{q0} & -i_{q0}^2\\ i_{d0}^2 & i_{d0}i_{q0}\end{bmatrix}, \mathbf{B_{Ic_v}}=\begin{bmatrix}-i_{q0}v_{d0} & -i_{q0}v_{q0}\\ i_{d0}v_{d0} & i_{d0}v_{q0}\end{bmatrix}\\ \mathbf{B_{Vc_i}}=\begin{bmatrix}-v_{cq0}i_{d0} & -v_{cq0}i_{q0}\\ v_{cd0}i_{d0} & v_{cd0}i_{q0}\end{bmatrix}, \mathbf{B_{Vc_v}}=\begin{bmatrix}-v_{cq0}v_{d0} & -v_{cq0}v_{q0}\\ v_{cd0}v_{d0} & v_{cd0}v_{q0}\end{bmatrix}\end{cases} \tag{4}$$

The detailed definitions of the variables can be found in [27].

Moreover, $G_P=\dfrac{1}{s\left(Js+D_p\right)}$ denotes the transfer function of the active power controller.

The impedance model $Z_{Pdq}$ with APCL dynamics is provided in [27], and the corresponding positive-sequence impedance $Z_{Pp}$ is plotted in Fig. 6. As seen, the inclusion of APCL results in a high impedance magnitude near the fundamental frequency, indicating current-source (or power-source) behavior. Moreover, a negative-resistance region is introduced around the fundamental frequency. This confirms that the APCL is the primary factor causing the impedance peak and the emergence of the negative-resistance region near the fundamental frequency in GFM converters.

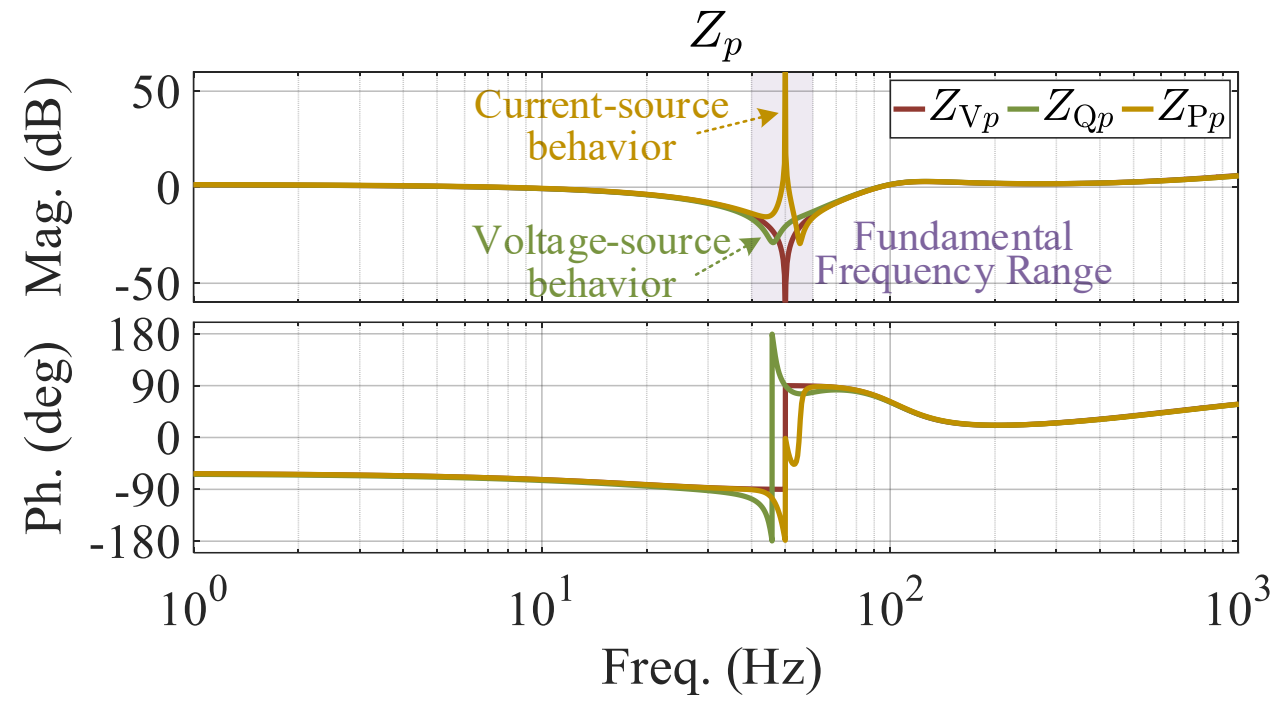

**Fig. 6.** Positive-sequence impedance of GFM converter with outer loops.

## III. APCL-Induced Impedance Peaks and Negative-Resistance Regions: Modeling and Analysis

The previous analysis shows that APCL is the primary cause of both the impedance peak and the negative-resistance region near the fundamental frequency. As shown in Fig. 7, its impact mainly stems from two factors: the gain of APCL ($G_P$) and the steady-state operating point (SSOP) matrix. The SSOP matrix characterizes the dynamics of the active-power perturbation $\hat{P}$, which is subsequently reflected in the impedance model through the APCL gain $G_P$. However, as shown in Fig. 7, the impedance matrix $Z_{Pdq}$ is highly complex [27], making it difficult to pinpoint the key contributors to the impedance peak. Therefore, this section derives an analytical expression of $Z_{Pdq}$ in terms of $G_P$ and the SSOP matrix to isolate the core factor causing the impedance peak and the negative-resistance region near the fundamental frequency.

### *A. SSOP-Based Analytical Modeling with APCL Dynamics*

As shown in (4), APCL introduces six SSOP matrices: $\mathbf{B_{Vo_i}}$, $\mathbf{B_{Vo_v}}$, $\mathbf{B_{Ic_i}}$, $\mathbf{B_{Ic_v}}$, $\mathbf{B_{Vc_i}}$, and $\mathbf{B_{Vc_v}}$. To determine which SSOP matrix contributes most significantly to the impedance peak of the GFM converter near the fundamental frequency, the contribution of each matrix is quantitatively evaluated. For this purpose, the following procedure is adopted:

1) The impedance peak obtained from the full-order model is first calculated and denoted as $Z_{peak}^{full}$.
2) The $i$-th SSOP matrix is then set to zero while keeping the remaining matrices unchanged, and the resulting impedance peak is obtained and denoted as $Z_{peak,\,S_i=0}$.
3) The contribution of the $i$-th SSOP matrix to the overall impedance peak is defined as the normalized difference between the two peaks, given by

$$C_i=\frac{Z_{peak}^{full}-Z_{peak,\,S_i=0}}{Z_{peak}^{full}}\times 100\% \tag{5}$$

where $C_i$ represents the relative contribution of the $i$-th SSOP matrix to the full-order impedance peak.

Based on the aforementioned contribution evaluation procedure, Fig. 8 presents the contributions of different SSOP matrices to the impedance peak under various power factors and operating points. It can be observed that, regardless of the operating conditions, $\mathbf{B_{Vo_v}}$ consistently exhibits the largest contribution to the impedance magnitude peak, far exceeding those of the other SSOP matrices. This indicates that $\mathbf{B}_{\text{Vo_v}}$ is the dominant SSOP matrix responsible for the impedance peak near the fundamental frequency. Therefore, without loss of generality, the following analysis considers the case where the GFM converter operates at unity power factor and rated power. In conjunction with Fig. 1, the PCC voltage is aligned with the *d*-axis. Consequently, the *q*-axis components of voltage and current in the SSOP matrices can be approximated as zero, while the *d*-axis components are rated values:

$$\begin{cases} \mathbf{B}_{\text{Vo_i}} \approx \begin{bmatrix} 0 & 0 \\ V_{\text{N}} I_{\text{N}} & 0 \end{bmatrix} & \mathbf{B}_{\text{Ic_i}} \approx \begin{bmatrix} 0 & 0 \\ I_{\text{N}}^2 & 0 \end{bmatrix} & \mathbf{B}_{\text{Vc_i}} \approx \begin{bmatrix} 0 & 0 \\ V_{\text{N}} I_{\text{N}} & 0 \end{bmatrix} \\ \mathbf{B}_{\text{Vo_v}} \approx \begin{bmatrix} 0 & 0 \\ V_{\text{N}}^2 & 0 \end{bmatrix} & \mathbf{B}_{\text{Ic_v}} \approx \begin{bmatrix} 0 & 0 \\ V_{\text{N}} I_{\text{N}} & 0 \end{bmatrix} & \mathbf{B}_{\text{Vc_v}} \approx \begin{bmatrix} 0 & 0 \\ V_{\text{N}}^2 & 0 \end{bmatrix} \end{cases} \tag{6}$$

where $V_{\text{N}}$ and $I_{\text{N}}$ are the rated voltage and current.

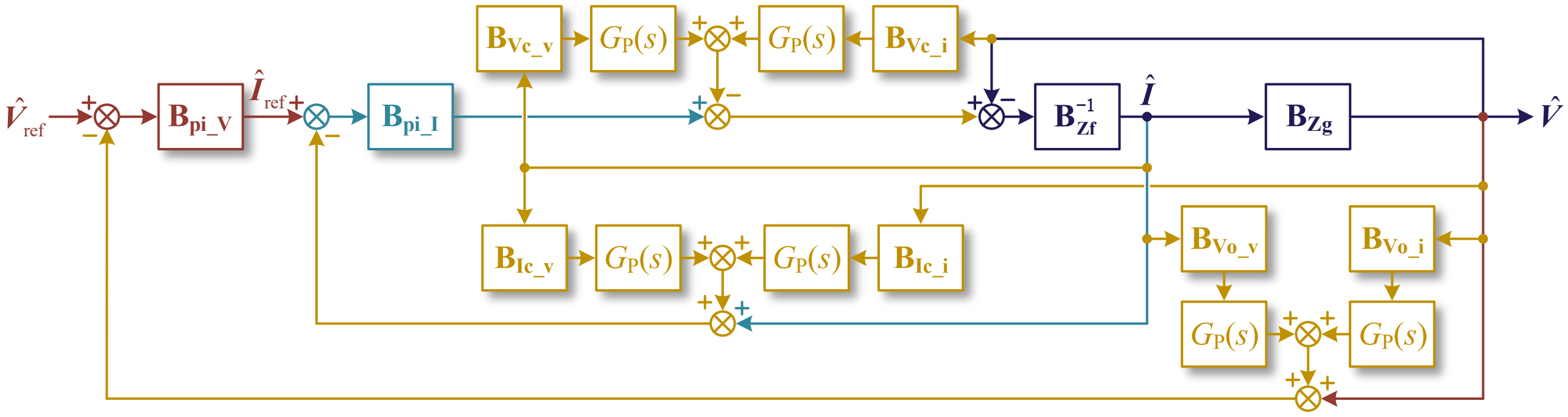

**Fig. 7.** Small-signal control block diagram of the GFM converter with APCL and inner loops.

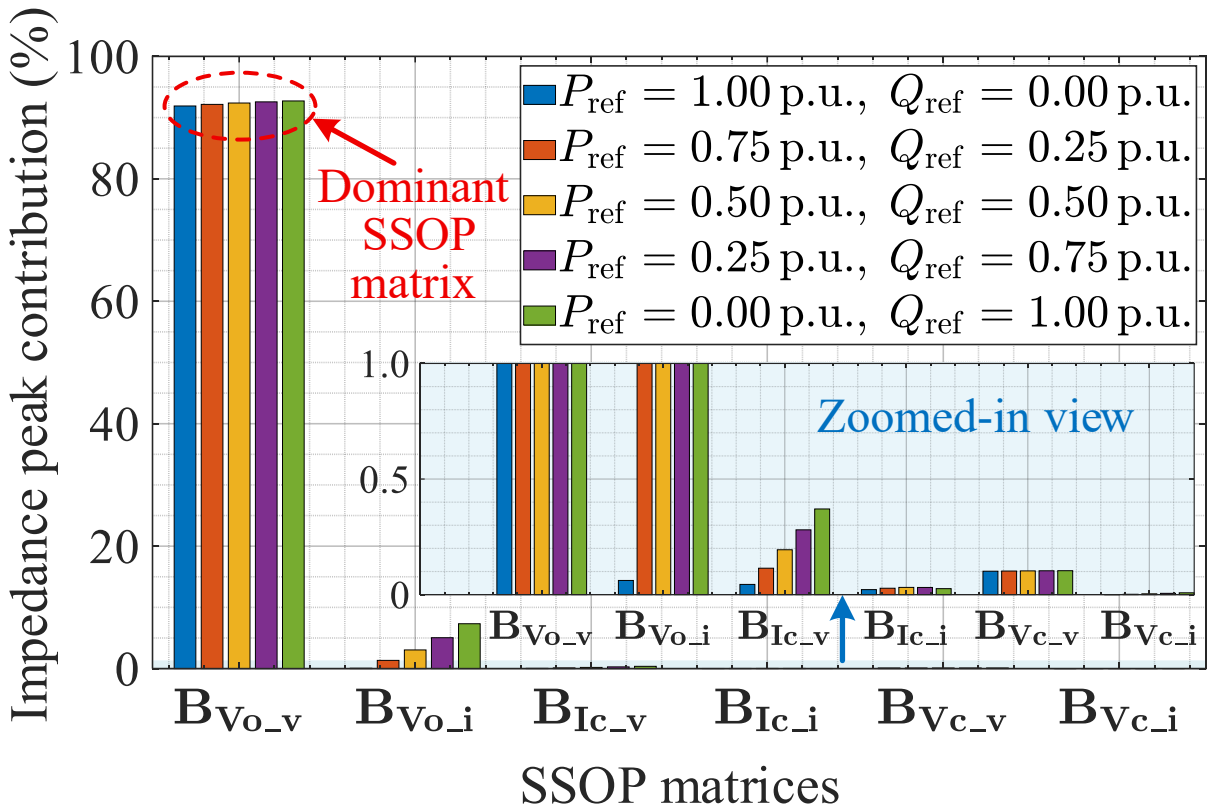

**Fig. 8.** Contributions of different SSOP matrices to the impedance peak of the GFM converter under various power factors and operating points.

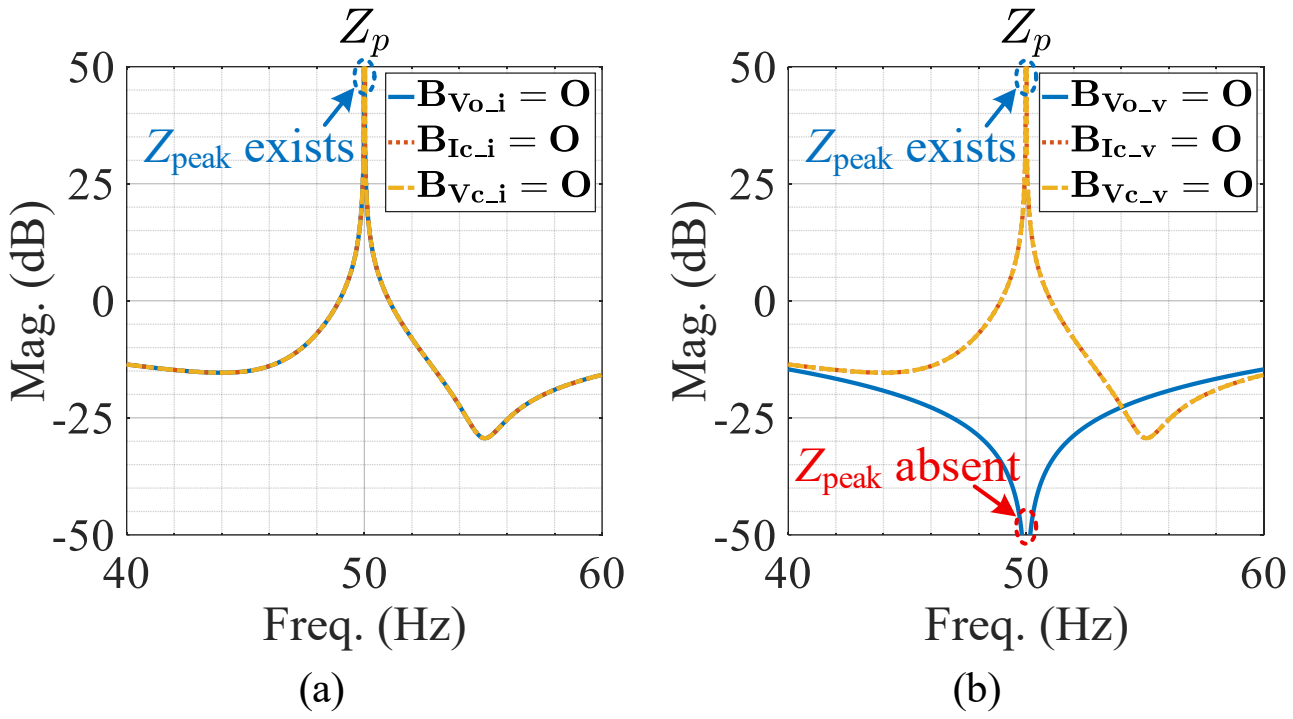

**Fig. 9.** Impact of individual SSOP matrices on fundamental-frequency impedance, where **O** denotes the zero matrix. (a) SSOP current matrices. (b) SSOP voltage matrices.

Following the above procedure, each of the six SSOP matrices is individually set to zero to examine its impact on the fundamental-frequency impedance, as illustrated in Fig. 9. The results show that the impedance magnitude decreases only when $\mathbf{B}_{\text{Vo_v}}=\mathbf{O}$, whereas in all other cases a peak still persists near the fundamental frequency. This further confirms that $\mathbf{B}_{\text{Vo_v}}$ is the dominant contributor to the impedance peak near the fundamental frequency. Therefore, by retaining only $\mathbf{B}_{\text{Vo_v}}$ and setting the remaining SSOP matrices to zero, a simplified model can be constructed to capture the dominant effect of the APCL dynamics. The corresponding small-signal control block diagram is shown in Fig. 10, from which a simplified impedance model $\boldsymbol{Z}_{\text{P}dq}^{\text{simp}}$ is derived:

$$\boldsymbol{Z}_{\text{P}dq}^{\text{simp}} = \begin{bmatrix} \dfrac{Z_{\text{f}} + G_{\text{I}}}{1 + G_{\text{V}} G_{\text{I}}} & 0 \\ \dfrac{V_{\text{N}}^2 G_{\text{P}} G_{\text{V}} G_{\text{I}}}{1 + G_{\text{V}} G_{\text{I}}} & \dfrac{Z_{\text{f}} + G_{\text{I}}}{1 + G_{\text{V}} G_{\text{I}}} \end{bmatrix} \tag{7}$$

From (7), it can be seen that the impact of APCL mainly appears in the coupling term ($q \rightarrow d$), while the decoupling terms ($d \rightarrow d$ and $q \rightarrow q$) remain in the same form as in (2). According to the transformation relationship between the scalar impedance and sequence impedance [26], the positive-sequence impedance of the simplified model is given by

$$\begin{aligned} Z_{\text{P}p}^{\text{simp}} &= \frac{1}{2}\left(\boldsymbol{Z}_{\text{P}dq,11}^{\text{simp}} + \boldsymbol{Z}_{\text{P}dq,22}^{\text{simp}}\right) + \frac{\text{j}}{2}\left(\boldsymbol{Z}_{\text{P}dq,21}^{\text{simp}} - \boldsymbol{Z}_{\text{P}dq,12}^{\text{simp}}\right) \\ &= \underbrace{\frac{Z_{\text{f}} + G_{\text{I}}}{1 + G_{\text{V}} G_{\text{I}}}}_{\boldsymbol{Z}_{\text{P}dq,11}^{\text{simp}}} + \frac{\text{j}}{2} \cdot \underbrace{\frac{V_{\text{N}}^2 G_{\text{P}} G_{\text{V}} G_{\text{I}}}{1 + G_{\text{V}} G_{\text{I}}}}_{\boldsymbol{Z}_{\text{P}dq,21}^{\text{simp}}} \end{aligned} \tag{8}$$

where the first term is identical to the impedance $Z_{\text{V}dq}$ obtained when the outer loop is neglected, as shown in (2). Based on the previous analysis, this term approaches zero near the fundamental frequency, and thus its magnitude also approaches zero. Expanding the second term $\boldsymbol{Z}_{\text{P}dq,21}^{\text{simp}}$ yields

$$\boldsymbol{Z}_{\text{P}dq,21}^{\text{simp}} = \frac{V_{\text{N}}^2}{s(Js + D_{\text{p}})} \frac{s^2 k_{\text{pI}} k_{\text{pV}} + s(k_{\text{pI}} k_{\text{iV}} + k_{\text{iI}} k_{\text{pV}}) + k_{\text{iI}} k_{\text{iV}}}{[s^2 (k_{\text{pI}} k_{\text{pV}} + 1) + s(k_{\text{pI}} k_{\text{iV}} + k_{\text{iI}} k_{\text{pV}}) + k_{\text{iI}} k_{\text{iV}}]} \tag{9}$$

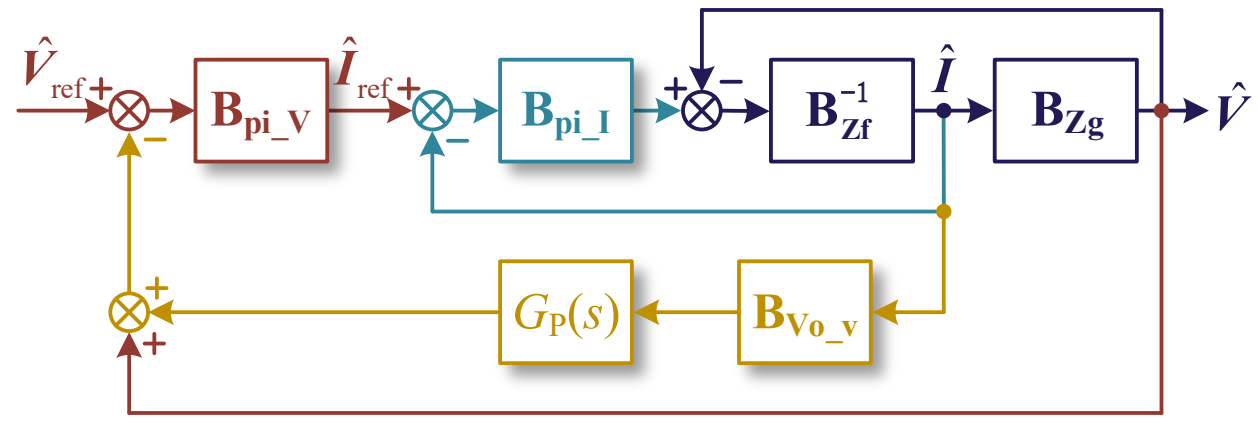

**Fig. 10.** Small-signal control block diagram with APCL and inner loops.

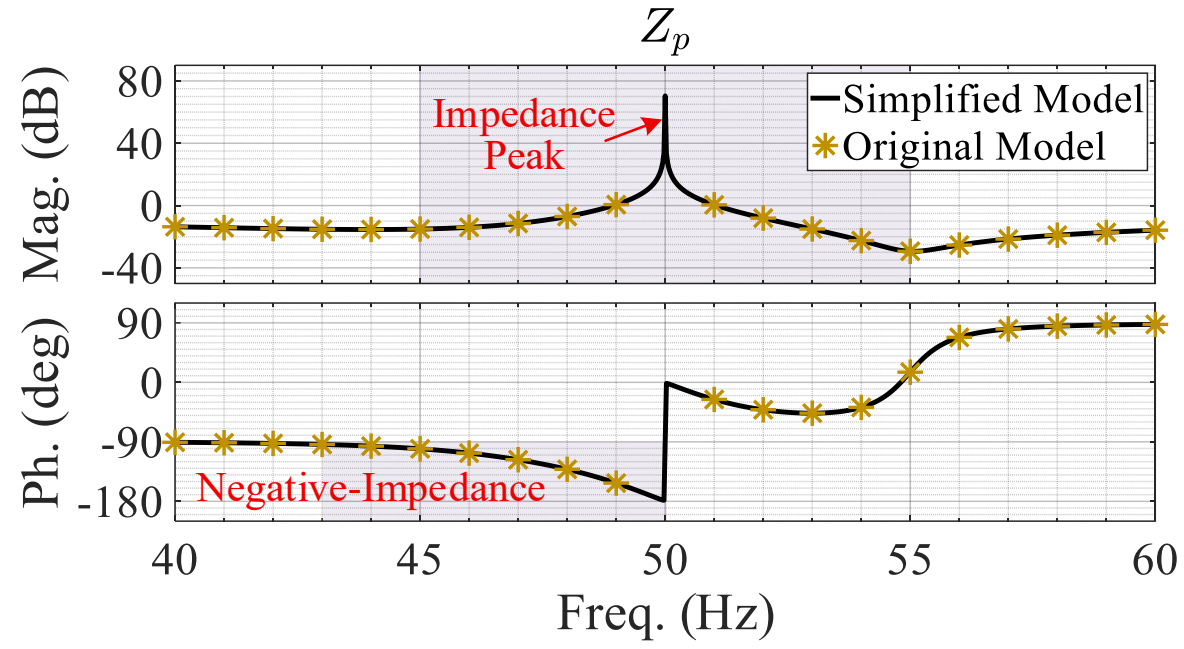

**Fig. 11.** Verification of the accuracy of the simplified impedance model.

It can be clearly observed that, due to the presence of the integrator, this term tends toward infinity near the fundamental frequency, and its magnitude behaves similarly. Consequently, the positive-sequence impedance of the simplified model in (8) increases in magnitude near the fundamental frequency and reaches a peak at the fundamental frequency.

To verify the correctness of the above theoretical analysis and the accuracy of the simplified model, Fig. 11 compares the frequency scan results with those of the original model. The results show that the impedance characteristics near the fundamental frequency are well preserved, including both the magnitude peak and the negative-resistance region, indicating that the simplified model successfully captures the key dynamics introduced by APCL. Therefore, the simplified model is validated and will be used in the subsequent analysis.

*B. Impedance Behavior under Different APCL Gain Designs*

The previous subsection identified that the SSOP matrix $\mathbf{B}_{\mathrm{Vo_v}}$, introduced by the APCL dynamics, particularly the active power perturbation $\hat{P}$, is the key factor determining whether an impedance peak appears near the fundamental frequency in GFM converters. Based on this finding, a simplified model of the GFM converter was established to capture the dynamics of the active power loop and to reflect the impedance characteristics around the fundamental frequency. This model provides analytical insight into why the GFM converter exhibits current-source behavior rather than voltage-source behavior near the fundamental frequency, as indicated by the magnitude responses in Figs. 6 and 11. It also explains why a positive resistance characteristic cannot be maintained over the entire frequency range, as reflected by the phase responses in Figs. 6 and 11. However, the specific frequency range in which the negative-resistance region appears, or equivalently the interval over which the source behavior is reshaped according to the magnitude response, has not yet been clearly determined. Therefore, this subsection further investigates the key factors that determine the frequency range of the impedance rise and the associated negative-resistance region.

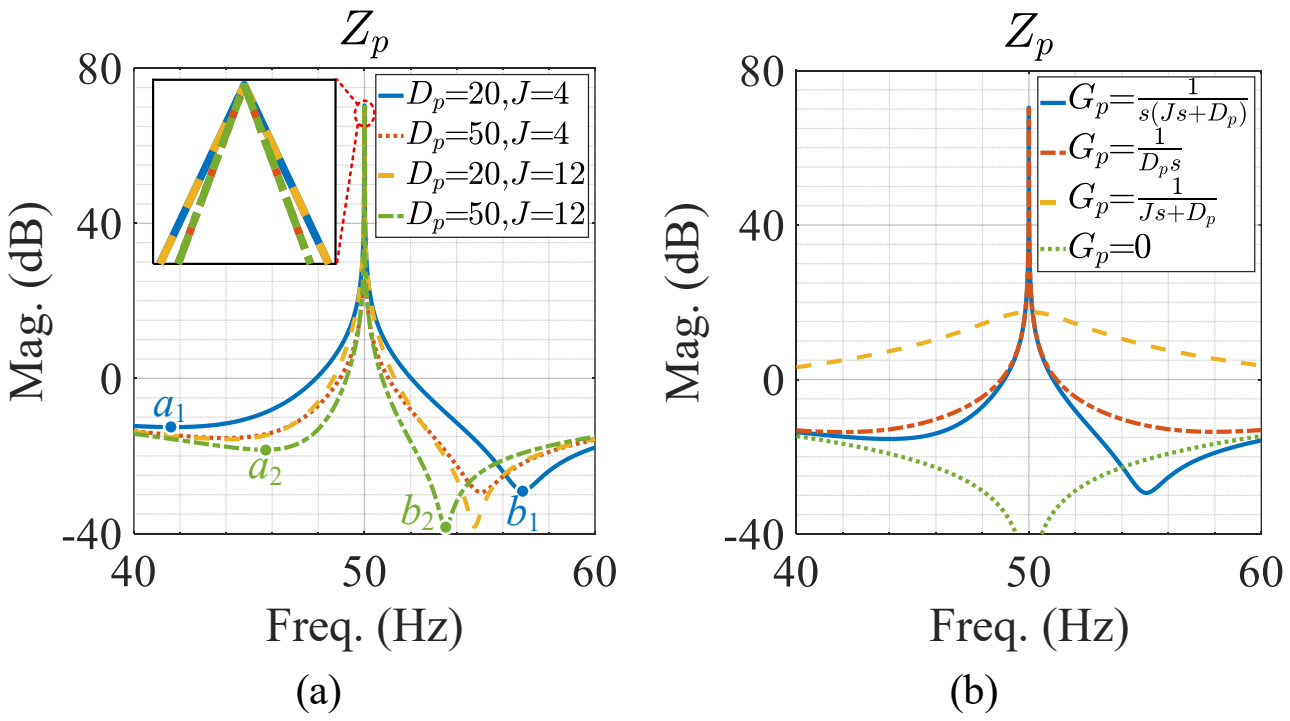

**Fig. 12.** Impact of APCL gain $G_P$ on fundamental-frequency impedance. (a) $G_P$ under varying parameters. (b) $G_P$ under different configurations.

As indicated by (7), the impact of APCL is reflected not only in the SSOP matrix but also in its gain $G_P$. The following analysis therefore focuses on how the APCL gain $G_P$ shapes the frequency range and magnitude of the impedance peak. The APCL gain $G_P$ consists of two parts: the control function that maps the power disturbance $\hat{P}$ to the frequency variation $\hat{\omega}$, given by $1/(Js+D_p)$, and the integrator that maps the frequency variation $\hat{\omega}$ to the phase angle variation $\hat{\theta}$, given by $1/s$. To examine their individual effects, Fig. 12 presents the GFM converter's positive-sequence impedance under different $G_P$ parameters and configurations.

As seen in Fig. 12(a), varying $D_p$ or $J$ (per unit value) in $G_P$ shifts the corner frequency of the impedance curve (e.g., points $a$ and $b$), where the magnitude begins to rise near the fundamental frequency, indicating the point at which the source behavior starts to transition. However, these parameters do not affect the peak value of the impedance. A larger $D_p$ or $J$ moves the corner frequency closer to the fundamental frequency, indicating a weaker impact of APCL on the converter behavior. Fig. 12(b) further shows that the integrator in $G_P$ is the key factor determining the peak magnitude, regardless of whether inertia is included in the GFM control. In summary, both the control function and the integrator in $G_P$ contribute to the impedance rise near the fundamental frequency, where the former sets the corner frequency of the magnitude increase and the latter determines the peak magnitude. It should be noted that although the APCL gain determines the magnitude and frequency range of the impedance peak near the fundamental frequency, this does not diminish the importance of the SSOP matrices introduced by APCL, particularly $\mathbf{B}_{\mathrm{Vo_v}}$, as discussed in the previous subsection. In fact, the presence of $\mathbf{B}_{\mathrm{Vo_v}}$ enables the dynamics of the active-power perturbation $\hat{P}$ to be mapped through the APCL gain $G_P$ into the final impedance model, thereby causing the GFM converter to exhibit current-source behavior and a negative-resistance region near the fundamental frequency.

Overall, the SSOP matrix $\mathbf{B}_{\mathrm{Vo_v}}$ introduced by APCL dynamics and the gain $G_P$, which contains the terms $1/(Js+D_p)$

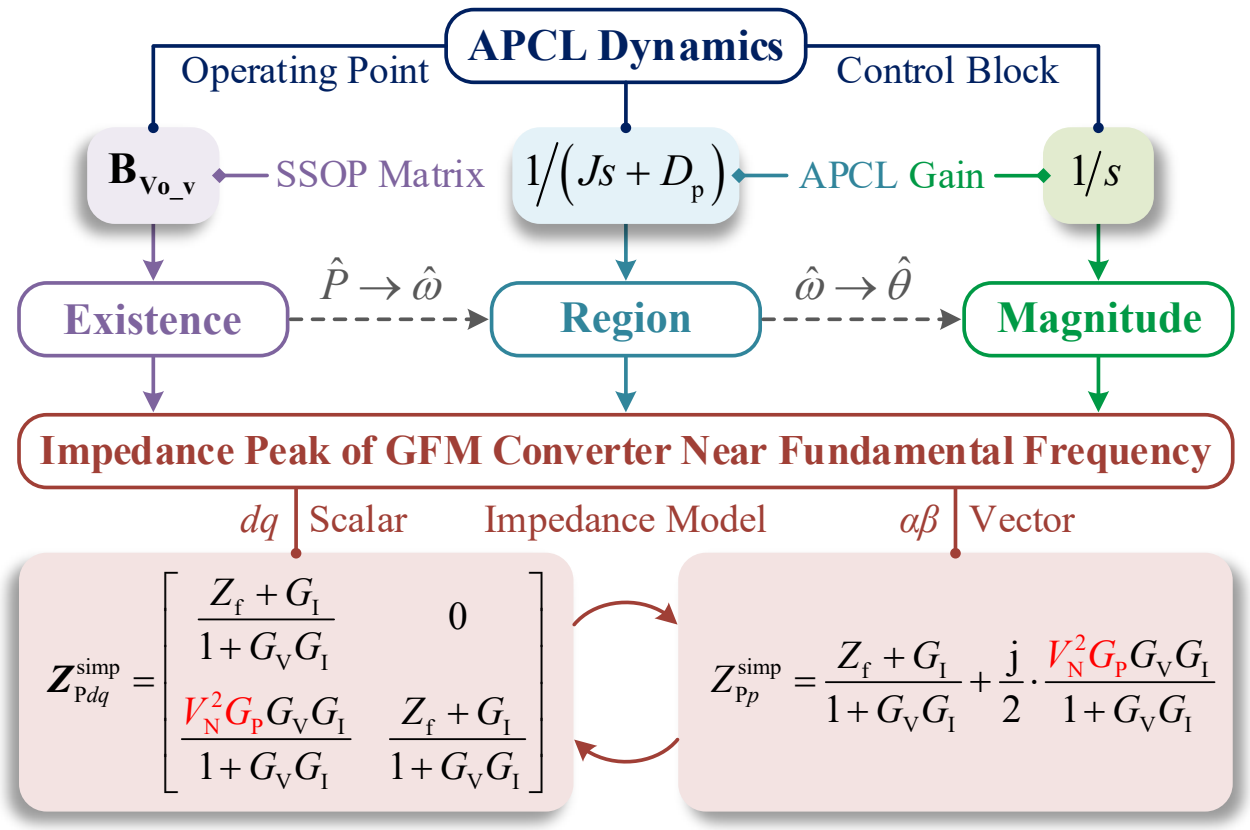

**Fig. 13.** Mechanism framework illustrating the impedance magnitude rise near the fundamental frequency caused by APCL dynamics in GFM converters.

and $1/s$, jointly govern the occurrence of the impedance peak near the fundamental frequency in GFM converters. Fig. 13 illustrates the mechanism framework through which APCL dynamics lead to the rise of impedance magnitude near the fundamental frequency. In this framework, the SSOP matrix determines whether the impedance peak exists, $1/(Js+D_p)$ determines the frequency range over which the magnitude increases, and $1/s$ determines the final peak magnitude. These three factors jointly govern the formation of this impedance characteristic from the perspectives of existence, frequency range, and magnitude, thereby revealing the fundamental mechanism behind the impedance peak of GFM converters near the fundamental frequency.

## IV. Exclusion Bandwidth Index for Impedance Characterization of GFM Converters

The analysis in Section III reveals that the APCL dynamics reshape the impedance characteristics of GFM converters near the fundamental frequency, leading to an impedance magnitude rise and a negative-impedance region. These phenomena complicate the assessment of voltage-source behavior and impedance characteristics. To provide a systematic way to address this issue, this section introduces a quantitative index for determining the exclusion bandwidth around the fundamental frequency and further investigates its sensitivity to APCL parameters and operating conditions.

### *A. Definition of the Exclusion Bandwidth Index*

GFM converters are widely expected to exhibit voltage-source behavior, with various metrics proposed to assess this attribute [9]. However, the above analysis shows that due to APCL, the converter exhibits current-source or power-source impedance characteristics near the fundamental frequency. This observation explains an engineering practice adopted in several technical standards and reports, where a narrow frequency band around the fundamental frequency is excluded when evaluating the impedance characteristics or voltage-source behavior of GFM converters. For example, the Finnish grid code specifies an exclusion band of ±3 Hz around the fundamental frequency [20], while similar requirements of ±4 Hz and ±5 Hz have been reported in the United States and China, respectively [21]-[23]. Nevertheless, these documents generally do not provide a clear theoretical basis for determining the width of the excluded frequency band.

To address this gap, this paper introduces a quantitative index for determining the exclusion bandwidth around the fundamental frequency. The index is based on the positive-sequence impedance magnitude curve $|Z_p(f)|$, where the corner frequencies $f_a$ and $f_b$ are identified as the points at which the slope of $|Z_p(f)|$ transitions from negative to positive around the fundamental frequency. Mathematically, these turning points satisfy:

$$\begin{cases} \left.\dfrac{d\left|Z_p(f)\right|}{df}\right|_{f=f_a^-} < 0, & \left.\dfrac{d\left|Z_p(f)\right|}{df}\right|_{f=f_a^+} > 0 \\ \left.\dfrac{d\left|Z_p(f)\right|}{df}\right|_{f=f_b^-} < 0, & \left.\dfrac{d\left|Z_p(f)\right|}{df}\right|_{f=f_b^+} > 0 \end{cases} \tag{10}$$

where $f_a<f_N$ and $f_b>f_N$, with $f_N$=50 Hz denotes the fundamental frequency. The left- and right-hand limits capture the slope immediately before and after each turning point, ensuring that the points are rigorously identified, such that the impedance magnitude rise begins at $f_a$ below $f_N$ and terminates at $f_b$ above $f_N$.

Based on these corner frequencies, the frequency intervals below and above the fundamental frequency are defined as

$$\Delta f_1 = f_N - f_a,\ \ \Delta f_2 = f_b - f_N \tag{11}$$

and the exclusion bandwidth is then quantified as

$$\Delta f = \max\{\Delta f_1, \Delta f_2\} \tag{12}$$

representing the minimum frequency range that should be excluded when evaluating the impedance characteristics or voltage-source behavior of the GFM converter.

To facilitate practical implementation in impedance testing and standardization, the exclusion bandwidth can be systematically determined according to the following procedure:

1) *Impedance Measurement:* Measure the positive-sequence impedance magnitude $|Z_p(f)|$ of the GFM converter across the frequency range of interest.
2) *Left Corner Frequency Identification:* Determine the left corner frequency $f_a$ by identifying the frequency below $f_N$ where the slope of $|Z_p(f)|$ changes from negative to positive (i.e., $d|Z_p|/df$ changes sign).
3) *Right Corner Frequency Identification:* Determine the right corner frequency $f_b$ by identifying the frequency above $f_N$ where the slope of $|Z_p(f)|$ changes from negative to positive.
4) *Interval Calculation:* Compute the intervals $\Delta f_1=f_N-f_a$ and $\Delta f_2=f_b-f_N$.
5) *Exclusion Bandwidth Determination:* Calculate the exclusion bandwidth as $\Delta f=\max\{\Delta f_1, \Delta f_2\}$.
6) *Band Exclusion:* Apply the exclusion band. For stationary-frame impedance, omit $f_N\pm\Delta f$; for rotating-frame impedance, omit $[0, \Delta f]$.

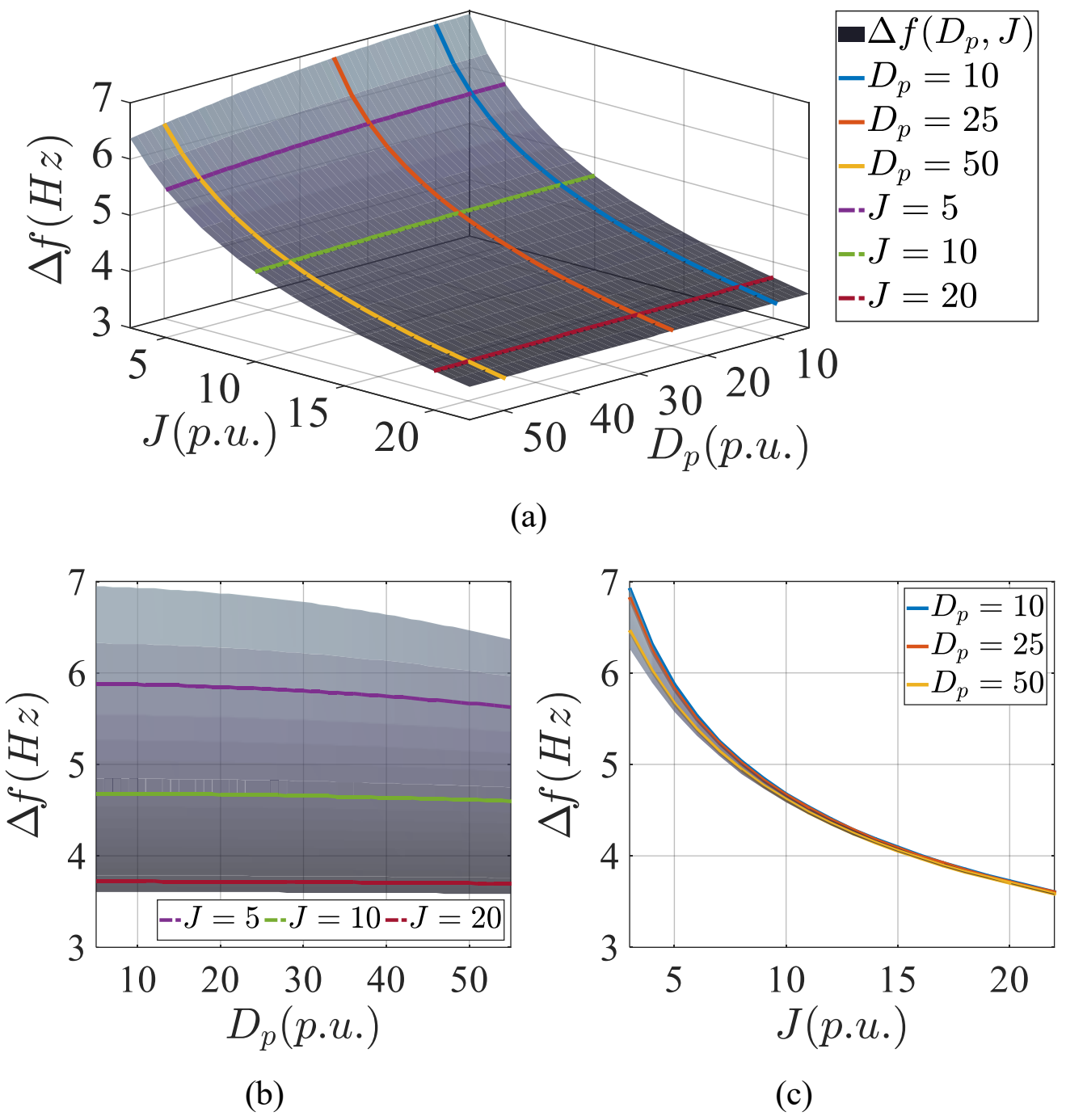

**Fig. 14.** Variation of $\Delta f$ with APCL parameters $D_p$ and $J$. (a) 3D surface of $\Delta f$ versus $D_p$ and $J$. (b) 2D projection of $\Delta f$ versus $D_p$ with fixed $J$. (c) 2D projection of $\Delta f$ versus $J$ with fixed $D_p$.

For example, considering the case in Fig. 12(a) with $D_p$=20 p.u. and $J$=4 p.u., the corner frequencies are identified as $f_a$=41.7 Hz and $f_b$=56.9 Hz. Accordingly, $\Delta f_1$=8.3 Hz and $\Delta f_2$=6.9 Hz are obtained, yielding an exclusion bandwidth of $\Delta f$=8.3 Hz according to (12). This example demonstrates how the proposed index provides a quantitative criterion for determining the frequency band that should be excluded when evaluating the voltage-source behavior of GFM converters, thereby offering a theoretical explanation for the exclusion practices adopted in existing grid codes.

It should be noted that although the frequency interval $[f_a,f_b]$ provides a more precise exclusion range, the proposed index $\Delta f$=max$\{\Delta f_1, \Delta f_2\}$ offers a practical and symmetric exclusion bandwidth, simplifying the implementation in impedance testing and standardization, while avoiding the overly conservative approach of directly using $[f_a, f_b]$.

*B. Sensitivity Analysis of the Proposed Index*

To further examine the effectiveness and robustness of the proposed index, its behavior under different internal parameters and external operating conditions is investigated. First, the influence of the APCL parameters on the proposed index $\Delta f$ is analyzed. Fig. 14 presents the relationship between $\Delta f$ and the APCL parameters $D_p$ and $J$, together with the corresponding two-dimensional projections showing the sensitivity to each parameter. As $D_p$ or $J$ increases, the effective gain of the APCL dynamics decreases, indicating a weaker influence of APCL on the converter impedance. Consequently, the exclusion bandwidth $\Delta f$ becomes narrower. This observation is consistent with the analytical results discussed earlier, as reflected in the impedance characteristics shown in Fig. 12. Moreover, $\Delta f$ is more sensitive to variations in $J$ than in $D_p$. Overall, these results confirm that the exclusion bandwidth $\Delta f$ is primarily governed by the APCL parameters.

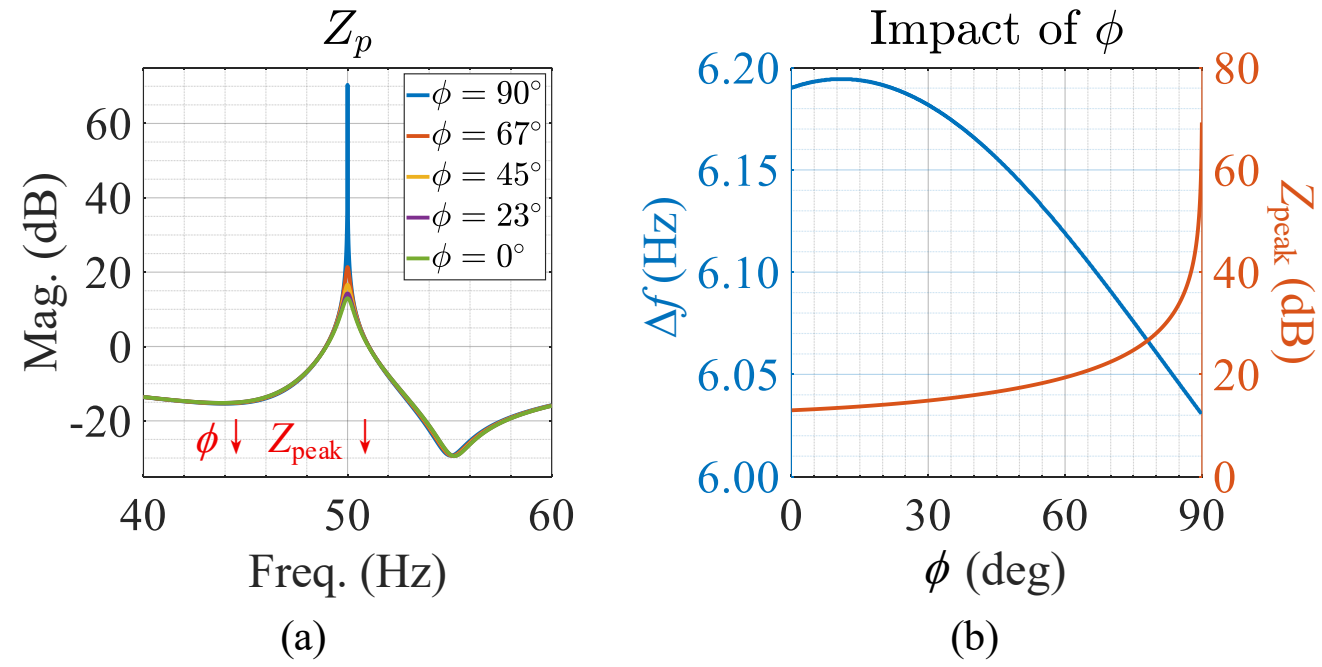

**Fig. 15.** Impedance characteristics of the GFM converter near the fundamental frequency under different power factors. (a) Impedance magnitude curves. (b) Variation of $\Delta f$ and $Z_{\text{peak}}$.

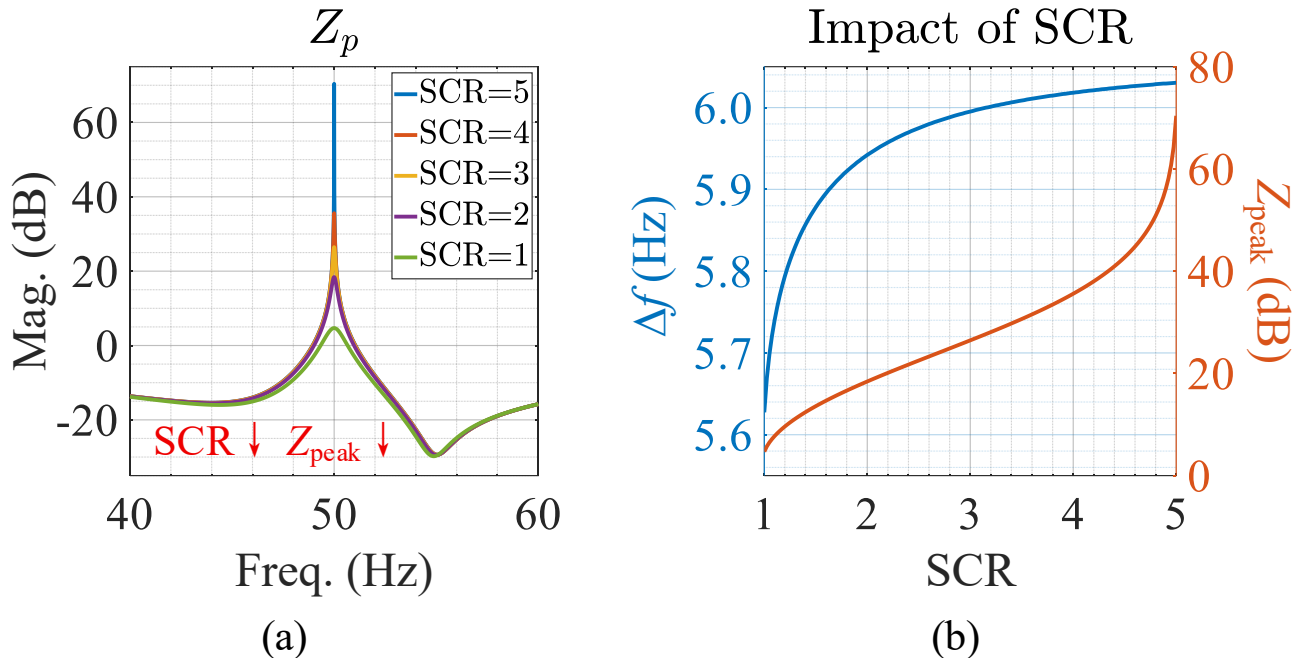

**Fig. 16.** Impedance characteristics of the GFM converter near the fundamental frequency under different grid strengths (SCR). (a) Impedance magnitude curves. (b) Variation of $\Delta f$ and $Z_{\text{peak}}$.

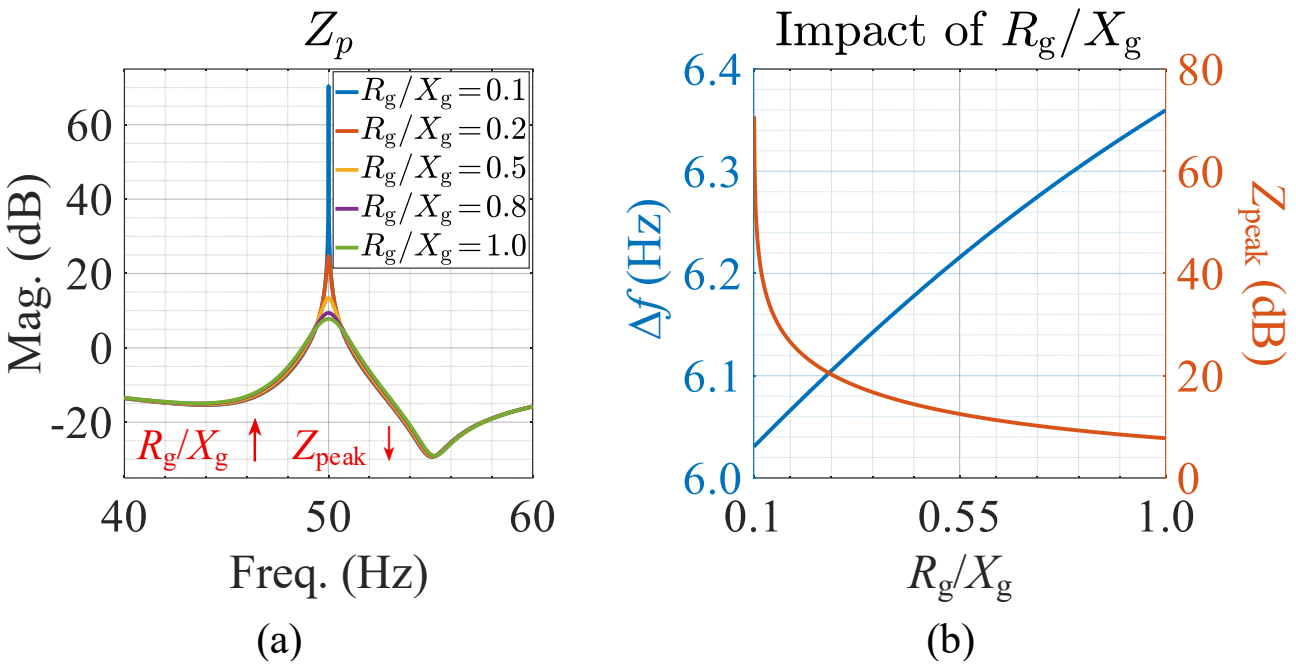

**Fig. 17.** Impedance characteristics of the GFM converter near the fundamental frequency under different line resistance-to-reactance ratios ($R_g/X_g$). (a) Impedance magnitude curves. (b) Variation of $\Delta f$ and $Z_{\text{peak}}$.

In addition, the robustness of the proposed index under different operating conditions is further investigated. Figs. 15–17 present the impedance characteristics of the GFM converter under different power factors, grid strengths, and line impedance ratios. Specifically, Fig. (a) in each case shows the impedance magnitude curves around the fundamental frequency, while Fig. (b) illustrates the variation of the exclusion bandwidth $\Delta f$ and the fundamental-frequency impedance peak $Z_{\text{peak}}$. As observed from the figures, the peak magnitude $Z_{\text{peak}}$ decreases as the power factor ($\phi$) decreases, the short-circuit ratio (SCR) becomes smaller, or the line resistance-to-reactance ratio ($R_g/X_g$) increases. Nevertheless, the impedance magnitude rise near the fundamental frequency consistently persists, although the peak values vary under different operating conditions. In contrast, the exclusion bandwidth $\Delta f$ exhibits only minor variations across these conditions, indicating that $\Delta f$ is largely insensitive to operating-point changes.

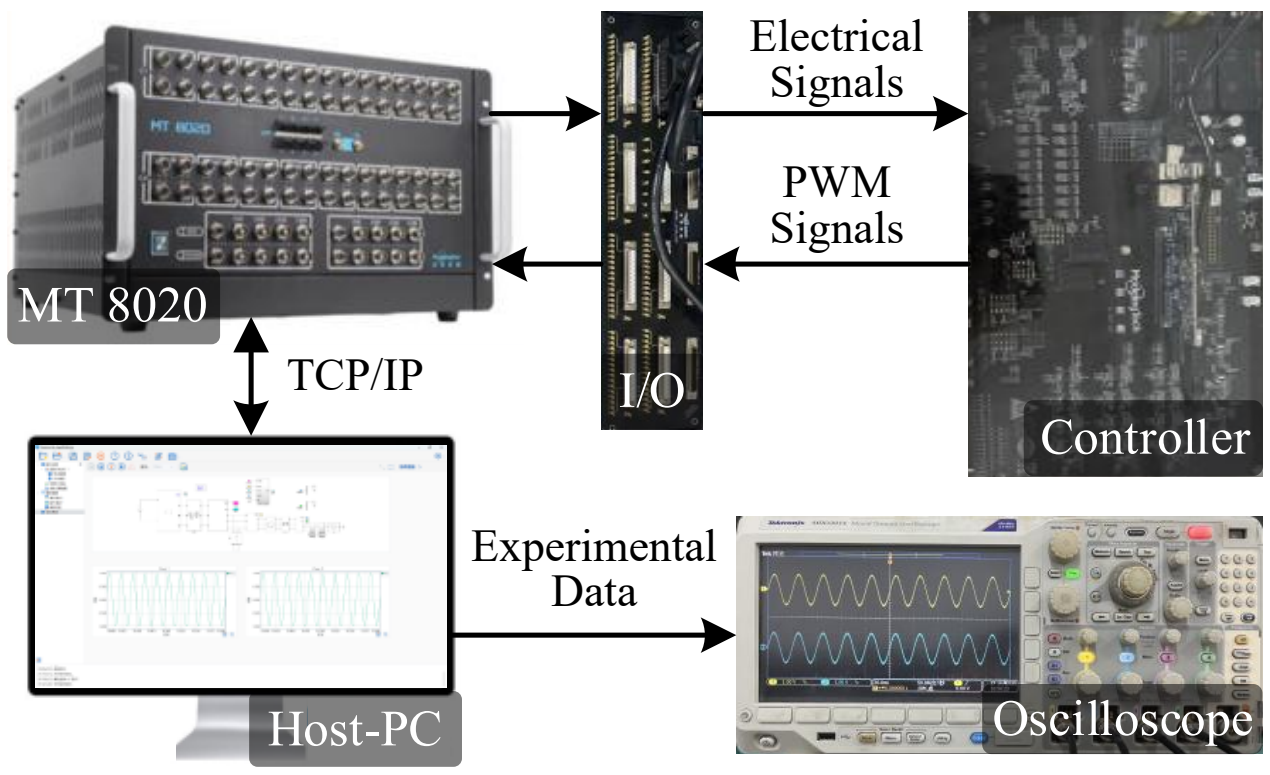

**Fig. 18.** Experimental verification platform for the GFM converter.

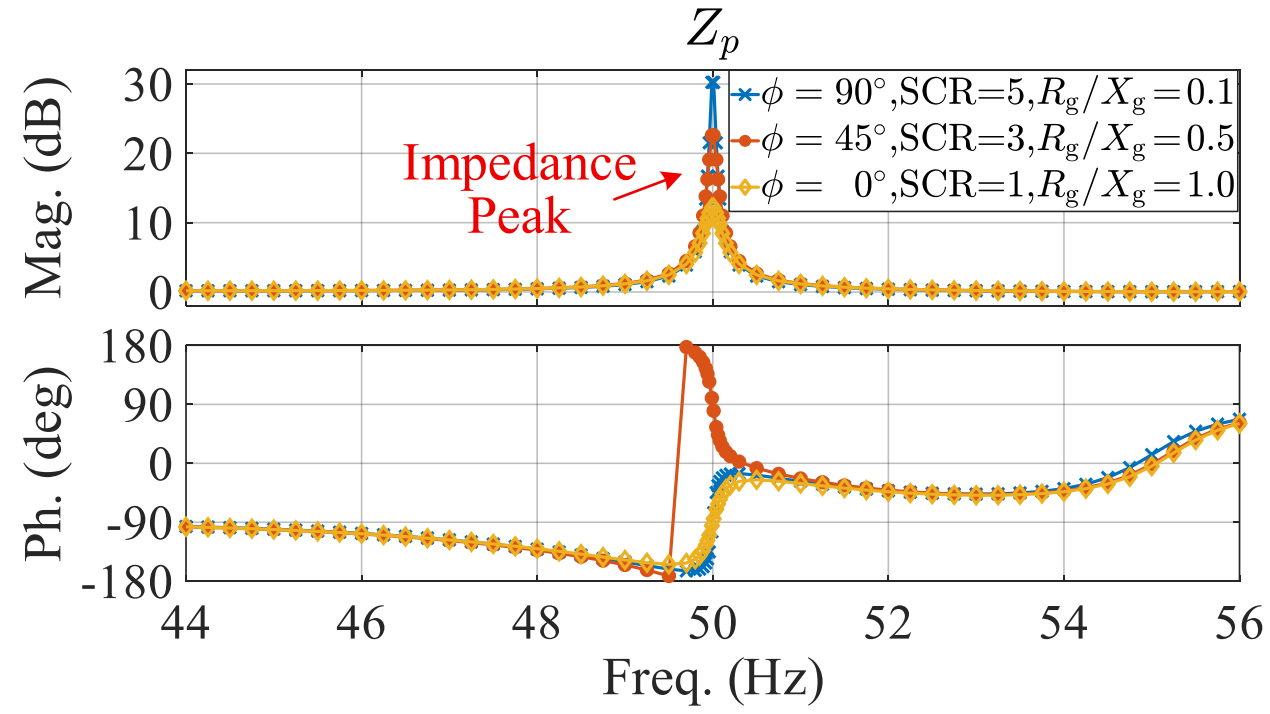

**Fig. 19.** Measured frequency-scan results of the GFM converter near 50 Hz.

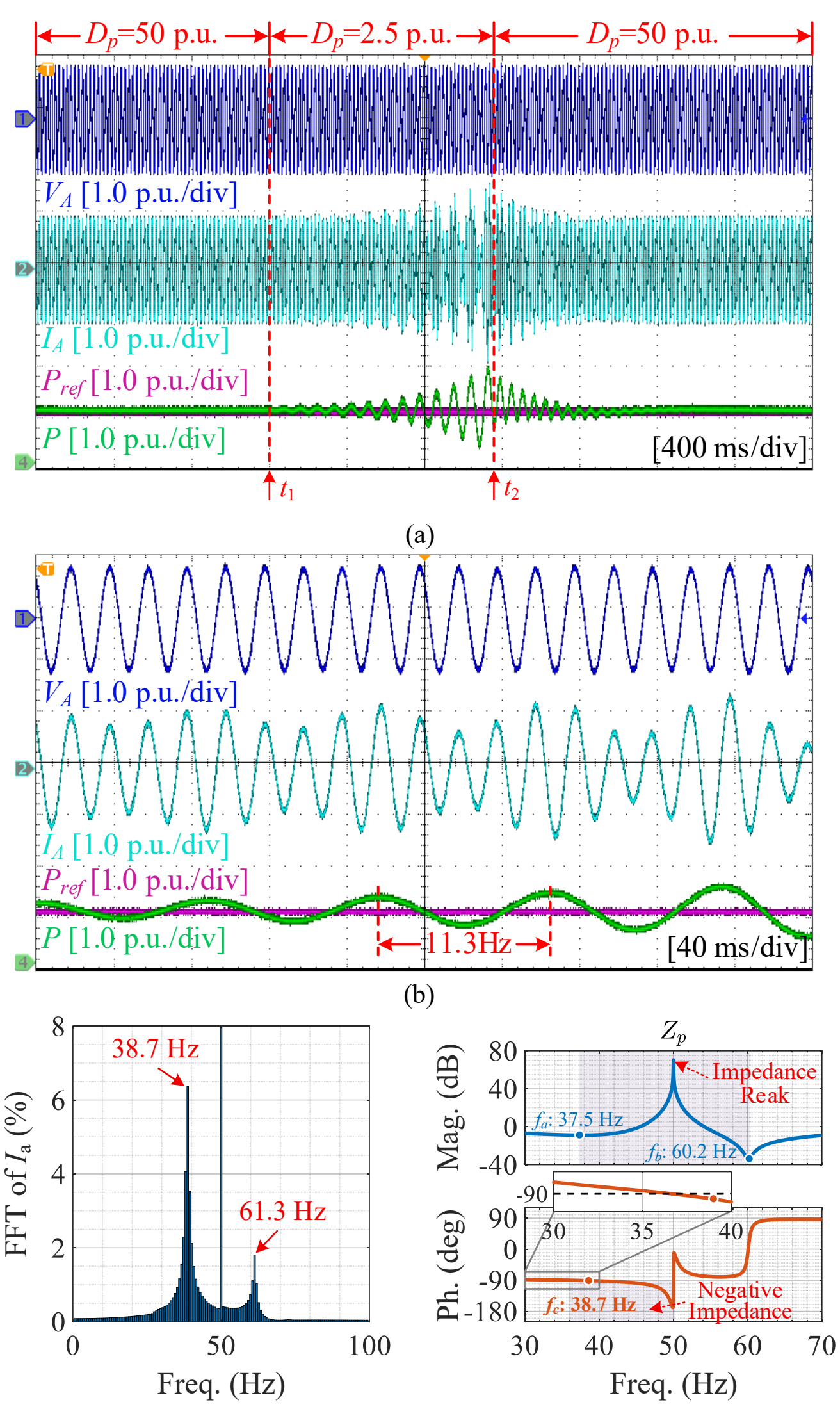

**Fig. 20.** Instability of the GFM converter under variations in the APCL damping coefficient $D_p$. (a) Overall waveforms. (b) Zoomed view. (c) FFT analysis. (d) Positive-sequence impedance with $D_p$=2.5 p.u.

It should be noted that the variations in these operating conditions essentially correspond to different steady-state operating points of the GFM converter. In the impedance model, such variations are reflected mainly through the SSOP matrix defined in (4). As discussed previously, the exclusion bandwidth $\Delta f$ is primarily determined by the APCL parameters and is much less sensitive to the SSOP matrix associated with different operating points. Thus, in practical applications, $\Delta f$ can be determined in advance from the APCL parameters $D_p$ and $J$, and subsequently adopted as a unified exclusion band for evaluating the voltage-source behavior and impedance characteristics of GFM converters. This index therefore provides a systematic basis for defining the exclusion bandwidth around the fundamental frequency and may serve as a reference for future impedance testing guidelines and grid-code development for GFM converters.

## V. Experimental Results

To validate the accuracy of the aforementioned theoretical analysis, experimental tests were carried out on a hardware-in-the-loop (HIL) platform. The power circuit models were implemented on a real-time simulator (StarSim 8020) with a fixed time step of 1 μs. The GFM-ESC control algorithm was executed on an MT1038 controller equipped with a DSP TMS320F28388D processor, operating at a control sampling period of 100 μs. The overall experimental setup is shown in Fig. 18, and the corresponding circuit topology and control parameters are summarized in Table I.

Fig. 19 presents the measured frequency-scan results of the GFM converter around 50 Hz. As observed, the impedance magnitude exhibits a pronounced rise near the fundamental frequency, with a distinct peak occurring at 50 Hz under all operating conditions. Moreover, the peak magnitude gradually decreases as the power factor decreases, the SCR becomes smaller, and the line resistance-to-reactance ratio increases, which is in full agreement with the theoretical results shown in Figs. 15–17. In terms of phase, a clear negative-impedance region is observed around 50 Hz, characterized by a prominent negative resistance-capacitance behavior. This negative-impedance characteristic may lead to stability issues when the GFM converter is interconnected with the power grid.

Furthermore, Fig. 20(a) shows the experimental waveforms of the GFM converter when the APCL damping coefficient $D_p$ varies. With $D_p$ initially set to 50 p.u., reducing it to 2.5 p.u. at $t_1$ induces pronounced oscillatory instability. The zoomed view in Fig. 20(b) shows that the active power oscillates at 11.3 Hz, while the FFT analysis in Fig. 20(c) indicates a component at 38.7 Hz in both grid voltage and current, and frequency coupling further introduces a 61.3 Hz component. This observation is consistent with the impedance characteristics in Fig. 20(d), where the positive-sequence impedance with $D_p$=2.5 p.u. exhibits a widened impedance rise region $\Delta f$ near the fundamental frequency, reflecting the strong shaping effect of APCL. In this region, the phase response indicates a negative resistance-capacitance characteristic, forming a negative-impedance interval around 38.7 Hz, which aggravates the interaction with inductive grid impedance and leads to low-frequency oscillations. When $D_p$ is increased back to 50 p.u. at $t_2$, the system returns to a stable state.

## VI. Conclusion

This paper identifies the mechanism behind the exclusion of a narrow band around the fundamental frequency in existing GFM converter standards that specify a positive-resistance range. Analytical modeling shows that the dynamics of the APCL, especially the mapping from power disturbance to the synchronous angle, inherently involve an integrative action. This feature drives the converter behavior from a voltage source toward a current source and thus prevents the formation of a positive-resistance characteristic near the fundamental frequency. It is further shown that the width of the negative-resistance region, which corresponds to the exclusion bandwidth defined in different grid codes, is directly determined by the APCL parameters, including inertia and damping. Larger parameter values lead to a narrower affected frequency range.

To enable consistent voltage-source assessment and support impedance standardization, a quantitative index is proposed based on the corner frequencies of the impedance magnitude peak, defining a symmetric exclusion interval around the fundamental frequency. This index provides a concise basis for voltage-source assessment and guides future impedance standardization. Future work will focus on integrating the proposed framework into system-level stability analysis and its incorporation into future grid-code development.